# Cooperative Adaptive Cruise Control for a Platoon of Connected and Autonomous Vehicles Considering Dynamic Information Flow Topology*

Siyuan Gong, Anye Zhou, Jian Wang, Tao Li and Srinivas Peeta

*Abstract*— Vehicle-to-vehicle communications can be unreliable as interference causes communication failures. Thereby, the information flow topology for a platoon of Connected Autonomous Vehicles (CAVs) can vary dynamically. This limits existing Cooperative Adaptive Cruise Control (CACC) strategies as most of them assume a fixed information flow topology (IFT). To address this problem, we introduce a CACC design that considers a dynamic information flow topology (CACC-DIFT) for CAV platoons. An adaptive Proportional-Derivative (PD) controller under a two-predecessor-following IFT is proposed to reduce the negative effects when communication failures occur. The PD controller parameters are determined to ensure the string stability of the platoon. Further, the designed controller also factors the performance of individual vehicles. Hence, when communication failure occurs, the system will switch to a certain type of CACC instead of degenerating to adaptive cruise control, which improves the control performance considerably. The effectiveness of the proposed CACC-DIFT is validated through numerical experiments based on NGSIM field data. Results indicate that the proposed CACC-DIFT design outperforms a CACC with a predetermined information flow topology.

## I. INTRODUCTION

Platoon control aims to minimize the speed differences among vehicles in a group while maintaining a stable and safe headway between adjacent vehicles [1]. It has significant potential to enhance traffic safety and highway capacity, and reduce fuel consumption [2]. Existing studies [1][3] show that autonomous vehicle (AV) platoons achieved through adaptive cruise control (ACC) can improve platoon safety and stability compared with platoons involving human drivers as AVs can execute platoon control strategies reliably and consistently. AV platoon performance can be further enhanced through cooperative adaptive cruise control (CACC) if all vehicles in the platoon are connected through vehicle-to-vehicle (V2V) communications that enables information exchange between vehicles. The communication capability in such connected autonomous vehicles (CAVs) also enables cooperative platoon control strategies to achieve system level objectives through coordination across all or some of the vehicles in the platoon.

The platoon control/CACC framework for CAVs typically consists of four components [4]: (i) node dynamics, which describe the dynamics of each vehicle in the platoon, (ii) information flow topology (IFT), which describes the configuration of V2V communication links among vehicles, (iii) distributed controller, which uses information from other vehicles in the platoon to devise control strategies, and (iv) formation geometry, which describes the desired headway between vehicles. Many studies have modeled the four components of a CAV platoon realistically in different scenarios. CAV dynamics are generally modeled using second-order vehicle dynamics models [5][6], third-order models or other nonlinear models [7]. Different controllers are designed to control the CAV platoon, including Proportional-Integral-Derivative (PID) controller [8][9], car-following model based controller [10], sliding mode controller [11], and model predictive controller [5][6][12] using a constant distance (CD) policy or a constant time headway (CTH) policy with some predetermined IFT (such as predecessor-following leader, two predecessor-following, two predecessor-following leader, and global communication).

Although a CAV platoon has several advantages over an AV platoon, the effectiveness of platoon control depends on the number of communication failures due to communication interference, especially when the ambient traffic contains a substantial number of connected vehicles. Thus, the evolving information flow topologies for a CAV platoon are critical to the performance of the platoon control, as they determine the amount of information being shared among individual vehicles through V2V communications. Several IFT configurations have been investigated in the literature to analyze platoon performance in terms of string stability, convergence rate, etc. For example, Zheng et.al [13] studied the influence of IFT for a homogeneous CAV platoon and introduced two methods to improve stability by carefully choosing the IFT. Gong et al. [5] illustrated analytically that the performance of platoon control in terms of string stability can be significantly improved by using a global communication topology, that is, each vehicle has the capability to send information to and receive information from all other vehicles in the platoon. However, most studies in CACC design assume a fixed and predetermined IFT. This assumption ignores the fact that the IFT can change dynamically due to communication failures [14][15][16]. The probability of a communication failure is proportional to the number of ongoing V2V communications occurring within the vehicle's communication range. Other factors, such as the dynamic CAV traffic flow density, also impact the number of ongoing communications within the communication range, and further impact communication failure. Hence, the IFT is less likely to be fixed in the real-world. In this context, a CACC based on a predetermined IFT may execute an erroneous control action, which diminishes

*This work is supported by the Center for Connected Automated Transportation (CCAT), the Region V University Transportation Center funded by the U.S. Department of Transportation, Award #69A3551747105.

S. Gong is with the NEXTRANS Center, Purdue University, West Lafayette, IN, 47907, USA (e-mail: gong131@purdue.edu).
A. Zhou is with School of Mechanical Engineering, Purdue University, West Lafayette, IN, 47907, USA, (e-mail: zhou843@purdue.edu).
J. Wang is with School of Civil Engineering, Purdue University, West Lafayette, IN, 47907, USA, (e-mail: wang2084@purdue.edu).
T. Li is Department of Computer Science, Purdue University, West Lafayette, IN, 47907, USA, (e-mail: taoli@purdue.edu).
S. Peeta is with School of Civil Engineering, Purdue University, West Lafayette, IN 47907 USA (phone: 765-496-9726; fax: 765-807-3123; e-mail: peeta@purdue.edu).

platoon performance related to mobility, stability and even safety.

To mitigate the negative effects of IFT dynamics, this paper introduces a CACC that considers a dynamic IFT for a CAV platoon, labeled the CACC-DIFT. An adaptive Proportional-Derivative (PD) controller under a two predecessor-following topology is proposed to reduce the negative effects of communication failures. When a communication failure occurs, instead of directly degrading into ACC, the CAV platoon will be controlled by a CACC obtained with the PD controller to maintain the desired control performance. The string stability of the designed PD controller is analyzed. The following four criteria are used to measure the performance of the controller: (i) vehicle stability (i.e. closed-loop stability for individual controller), (ii) spacing error rectification, (iii) noise mitigation, and (iv) the ability to avoid actuator saturation. The effectiveness of the proposed CACC-DIFT is validated through simulation experiments based on NGSIM field data [17]. The study results indicate that the proposed design outperforms the CACC with a predetermined IFT.

The remainder of the paper is organized as follows. Section II briefly introduces the CACC based on the two predecessor-following topology and the degeneration that occurs due to the IFT dynamics. The proposed CACC-DIFT controller is formulated in Section III. Section IV discusses the parameters in CACC-DIFT by considering the aforementioned four criteria and string stability. The proposed design is validated using numerical experiments presented in Section V. Concluding comments are provided in Section VI.

## II. PROBLEM FORMULATION

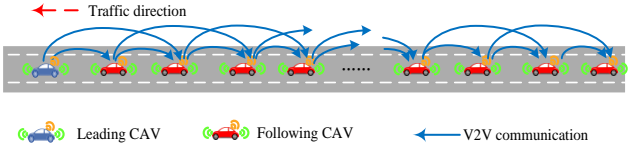

Figure 1. The information topology for CAV platooon with two predecessor-following information topology

As shown in Fig.1, this study considers a CAV platoon where information of each CAV is intended to be delivered to the two vehicles immediately following it through V2V communications (labeled as two predecessor-following IFT). Each CAV $i$ is also able to the detect the state of its immediate predecessor vehicle's (vehicle $i-1$) kinematic state (i.e. location $x_{i-1}$ and speed $v_{i-1}$) through on-board sensors such as radar, Lidar and camera (see Fig. 2). The acceleration rates of its two predecessors (i.e. $\ddot{x}_{i-1}$ and $\ddot{x}_{i-2}$) are obtained using V2V communications.

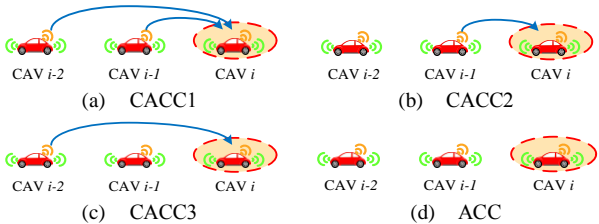

Figure 2. The proposed CACC-DIFT

Due to communication interference, V2V communications between two arbitrary CAVs can fail [15]. Then, the expected IFT (i.e. CACC1 in Fig. 2(a)) may degenerate into the three scenarios shown in Figs 2(b)-2(d). Figs. 2(b) and 2(c) show the cases when one communication link fails. In these cases, the CAV $i$ can detect the kinematic state of its immediate predecessor $i-1$, and one predecessor vehicle's acceleration rate through V2V communications. When both communication links fail (Fig. 2(d)), the acceleration rates of its two predecessor vehicles will not be known to CAV $i$. Then, the CACC will reduce to a traditional ACC to update the acceleration based on the relative spacing and speed between CAVs $i$ and $i-1$.

Based on these four scenarios, this paper seeks to develop four sets of controllers which will be switched adaptively based on the dynamic IFTs.

## III. DESIGN OF CACC-DIFT

This study assumes all CAVs in the platoon to be identical, forming a homogeneous vehicle string. The control schematic of vehicle $i$ in the platoon is described in Fig 3.

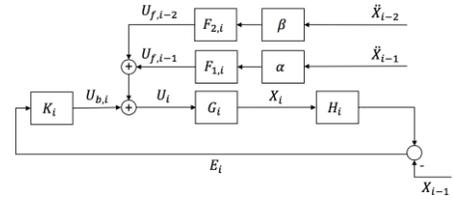

Figure 3. Block diagram of control schematic

In Fig. 3, $U_i$ represents the control command, which consists of control feedback $U_{b,i}$ from the error $E_i$ and two extra feedforward terms $U_{f,i-1}$ and $U_{f,i-2}$ from the acceleration rates $\ddot{X}_{i-1}$ and $\ddot{X}_{i-2}$, respectively. $X_i$ is the position output, $X_{i-1}$ is the feedback position information from the immediate predecessor. $K_i$ is the feedback controller which generates a control command to rectify the error. $G_i$ represents the ideal longitudinal vehicle dynamics. $H_i$ denotes the spacing policy (i.e. such CD and CTH), and $F_{1,i}$ and $F_{2,i}$ are feedforward filters to process the acceleration information from the corresponding predecessor vehicles. $\alpha$ and $\beta$ are indicators for the success of V2V communications ($\alpha$ and $\beta$ are equal 1 to for a successful communication between the CAV and the corresponding predecessor vehicles, and 0 otherwise). These terms are explained in detail hereafter.

### A. Vehicle dynamics

We ignore the air drag, rolling resistance and actuator delay in the vehicle dynamics model. The linearized state-space representation of the idealized longitudinal vehicle dynamics can be represented as:

$$\dot{x}_i(t) = v_i(t) \quad (1)$$
$$\dot{v}_i(t) = u_i(t) \quad (2)$$

where $x_i(t)$, $v_i(t)$, $u_i(t)$ are the absolute position, velocity and acceleration of vehicle $i$ at time $t$, respectively.

To analyze stability performance, the modeling and analysis are performed in the frequency domain. Hence, the idealized longitudinal vehicle dynamics in the Laplace domain can be described by using a transfer function:

$$G_i(s) = \frac{X_i(s)}{U_i(s)} = \frac{1}{s^2} \quad (3)$$

where the input $U_i(s)$ denotes the acceleration of vehicle $i$ and the output $X_i(s)$ denotes the absolute position of vehicle $i$ in the Laplace domain.

## B. Spacing policy

For simplicity, a CTH policy is used to model the desired relative distance between adjacent vehicles as follows:
$$d_i(t) = L + h_d \dot{x}_i(t) \quad (4)$$
where $d_i(t)$ is the desired relative distance between vehicle $i$ and vehicle $i-1$, $L$ is the constant standstill distance between the two vehicles, $\dot{x}_i(t)$ is the velocity of vehicle $i$ and $h_d$ is the desired time headway.

According to Eq. (4), the spacing error is
$$e_i(t) = x_{i-1}(t) - x_i(t) - d_i(t)$$
$$= x_{i-1}(t) - x_i(t) - (L + h_d \dot{x}_i(t)) \quad (5)$$
In Laplace domain, the spacing error can be expressed equivalently as:
$$E(s) = X_{i-1}(s) - H_i(s) X_i(s) \quad (6)$$
where $H_i(s)$ is the CTH spacing policy formulated as
$$H_i(s) = 1 + h_d s \quad (7)$$

## C. Acceleration feedforward

In CACC, the acceleration rate of the two immediate predecessors can be obtained through V2V communications. The acceleration rate data is used generate a feedforward control signal $U_{f,i-1}$ though a feedforward filter $F_i(s)$. To eliminate the spacing error between adjacent vehicles, the feedforward filter is designed based on a zero-error condition [9], where the relationship between tracking error $E_i(s)$ and feedforward acceleration $X_{i-1}(s)$ of the predecessor $i-1$ in Laplace domain is formulated as:
$$E_i(s) = \frac{1 - H_i(s) G_i(s) F_i(s) s^2}{1 + H_i(s) G_i(s) K_i(s)} X_{i-1}(s) \quad (8)$$

To satisfy the zero-error condition, the numerator of the right side in Equation (8) should be zero. By combining Equation (3), we have
$$F_i(s) = (H_i(s) G_i(s) s^2)^{-1} = \frac{1}{H_i(s)} \quad (9)$$

We apply the same kind of feedforward filter to the acceleration information from the second predecessor $\ddot{X}_{i-2}(s)$ as well. Hence, we can define the feedforward filters $F_{1,i}(s)$ for acceleration from the first predecessor and $F_{2,i}(s)$ for acceleration from the second predecessor as:
$$F_{1,i}(s) = F_{2,i}(s) = \frac{1}{H_i(s)} = \frac{1}{1 + h_d s} \quad (10)$$

## D. Control command

As illustrated in Fig.3, our control command consists of a feedback term and two feedforward terms:
$$U_i(s) = U_{b,i}(s) + U_{f,i-1}(s) + U_{f,i-2}(s) \quad (11)$$

Recall that the feedback term $U_{b,i}(s)$ uses spacing error to stabilize the closed-loop system while feedforward terms $U_{f,i-1}(s)$ and $U_{f,i-2}(s)$ use acceleration rate from predecessors to minimize the spacing error. Next, we determine the three terms in detail analytically.

The feedback term $U_{b,i}(s)$ and the corresponding PD feedback controller are defined as [9]:
$$U_{b,i}(s) = K_i(s) E_i(s) \quad (12)$$
$$K_i(s) = \omega_{K,i}(\omega_{K,i} + s) \quad (13)$$
where $\omega_{K,i}$ is the cut-off frequency of the PD controller. It has a strong impact on the string stability of the platoon, and will be determined analytically in Section IV. $E_i(s)$ is the spacing error in the Laplace domain.

The first feedforward term $U_{f,i-1}(s)$ indicates that the acceleration rate information of vehicle $i-1$ is sent to $i$:
$$U_{f,i-1}(s) = \alpha F_{1,i}(s) s^2 X_{i-1}(s) \quad (14)$$

The second feedforward term $U_{f,i-2}(s)$ indicates that the acceleration rate information of vehicle $i-2$ is sent to $i$:
$$U_{f,i-2}(s) = \beta F_{2,i}(s) s^2 X_{i-2}(s) \quad (15)$$

Note that according to the two predecessor-following IFT, the second CAV in the platoon can only receive acceleration information from the leading CAV, that is, the feedforward term of vehicle 1 only includes $U_{f,0}(s)$.

The overall control command is obtained by summing up equations (12), (14) and (15). Through inverse Laplace transformation, the expression for the control command can be formulated as
$$u_i(t) = \omega_{K,i}^2 e_i(t) + \omega_{K,i} \dot{e}_i(t)$$
$$+ \alpha F_{1,i}(t) \ddot{x}_{i-1}(t) + \beta F_{2,i}(t) \ddot{x}_{i-2}(t) \quad (16)$$

To implement the spatio-temporal dynamics of control command, the continuous control command in equation (16) will be discretized. At time step $k$, the control command is
$$u_i(k) = K_{p,i} e(k) + K_{d,i} \dot{e}(k) + u_{f,i-1}(k) + u_{f,i-2}(k) \quad (17)$$
where,
$$e(k) = X_{i-1}(k-1) - X_i(k-1) - h_d V_i(k-1) - L \quad (18)$$
$$\dot{e}(k) = V_{i-1}(k-1) - V_i(k-1) - h_d u_i(k-1) \quad (19)$$
$$u_{f,i-1}(k) = \frac{T}{h_d}\left(-u_{ff1,i}(k-1) + \alpha u_{i-1}(k-1)\right) \quad (20)$$
$$u_{f,i-2}(k) = \frac{T}{h_d}\left(-u_{ff2,i}(k-1) + \beta u_{i-2}(k-1)\right) \quad (21)$$

## IV. STRING STABILITY AND DETERMINATION OF PARAMETERS

There are two parameters in the designed system that significantly impact the platoon performance: the time headway $h_d$ and cut-off frequency $\omega_{K,i}$. This section seeks to analyze the inputs for the two parameters to improve the performance of individual vehicles while ensuring the string stability of the platoon. However, due to space constraints for the paper, the design process cannot be articulated in detail. Hence, we summarize the factors and key findings from the individual vehicle performance perspective and then focus on analyzing string stability.

### A. Performance of individual vehicle

The designed controls seek to maximize the performance of individual vehicle related to the following four criteria:

*1) Stability of individual vehicle.* The dynamics of an individual vehicle are stable if the states of this vehicle can converge to its equilibrium point asymptotically.

*2) Spacing error.* The spacing error should be minimized by the controllers to ensure the stability of platoon. Note that the steady-state error is proportional to $\omega_{K,i}^{-1}$; a lower bound should be set for $\omega_{K,i}$.

*3) Noise mitigation.* The high frequency measurement noise effect in the control process that may cause control signal chatter should be minimized. This control goal entails an upper bound of $\omega_{K,i}, h_d$ based on the noise mitigation level the system requires.

*4) Avoiding actuator saturation.* It indicates that the closed-loop bandwidth should not be larger than the bandwidth of the vehicle dynamics model (such that the control command will not exceed the vehicle acceleration/deceleration capability), setting up an upper bound for $\omega_{K,i} h_d$ as well.

## B. String stability

The string stability transfer function is defined as a measure of the signal amplification upstream from a platoon. In this paper, the head-to-tail string stability will be analyzed to ensure that the oscillations in the upstream traffic will be dampened when they reach the tail of the CAV platoon. The transfer function is:

$$SS_{X,i} = \frac{X_i}{X_0} \quad (22)$$

To ensure a head-to-tail string stability condition, we have

$$\|SS_{X,i}(j\omega)\|_\infty = \left\|\frac{X_i(j\omega)}{X_0(j\omega)}\right\|_\infty \leq 1 \quad (23)$$

where the $\infty$-norm indicates that the magnitude of $|SS_{X,i}(j\omega)| \leq 1$ for all $\omega, j = \sqrt{-1}$.

According to Equations (3), (6), (8), (9), (12), (14), and (15), the absolute position of all vehicles in the platoon can be described as:

$$\bar{X}_n = M\bar{X}_{n-1} \quad (24)$$

where $\bar{X}_n = (X_1, X_2, \ldots, X_N)^T$, $\bar{X}_{n-1} = (X_0, X_1, \ldots, X_{N-1})^T$,

$$M = \begin{pmatrix} g_{11} & 0 & \cdots & 0 & 0 \\ g_{21} & g_{22} & \cdots & 0 & 0 \\ \vdots & \vdots & \ddots & \vdots & \vdots \\ 0 & 0 & \cdots & g_{n-1,n-1} & 0 \\ 0 & 0 & \cdots & g_{n,n-1} & g_{n,n} \end{pmatrix}$$

$$g_{i,i} = \alpha\Lambda_{f,i-1} + \Lambda_{b,i-1} \quad (25)$$
$$g_{i,i-1} = \beta\Lambda_{f,i-2} \quad (26)$$

Also, the position of vehicle $i$ ($i>1$) can be described as:

$$X_i = \beta\Lambda_{f,i-2}X_{i-2} + (\alpha\Lambda_{f,i-1} + \Lambda_{b,i-1})X_{i-1} \quad (27)$$

where $\Lambda_{f,i-2} = \frac{G_i F_{2,i} s^2}{1+G_i K_i H_i}$ is the transfer function between position of vehicle $i-2$ and vehicle $i$ with feedforward term $U_{f,i-2}(s)$; $\Lambda_{f,i-1} = \frac{G_i F_{1,i} s^2}{1+G_i K_i H_i}$, represents the transfer function between vehicle $i-1$ and vehicle $i$ with feedforward term $U_{f,i-1}(s)$; $\Lambda_{b,i-1} = \frac{G_i K_i}{1+G_i K_i H_i}$, represents the transfer function between the position of vehicle $i-1$ and vehicle $i$ with the feedback term $U_{b,i}(s)$.

Based on the four possible communication scenarios described in Fig. 2, we will analyze the corresponding feasible region for the time headway $h_d$ and cut-off frequency $\omega_{K,i}$ to ensure string stability.

*1) CACC1 Case:* When $\alpha = 1$ and $\beta = 1$, the head-to-tail string stability transfer function is

$$SS_{X,i} = \Lambda_{f,i-2}\frac{X_{i-2}}{X_0} + (\Lambda_{f,i-1} + \Lambda_{b,i-1})\frac{X_{i-1}}{X_0} \quad (28)$$

Note that to analyze string stability based on this equation (28) is complex as it is a high order transfer function. To address this problem, we only consider the worst-case scenario in equation (8) where the value of both $X_{i-2}/X_0$ and $X_{i-1}/X_0$ are equal to 1 (both are head-to-tail marginally string stable[1]). This enables us to find a more conservative, stable region of the two parameters to ensure string stability. When $X_{i-2}/X_0 = X_{i-1}/X_0 = 1$, equation (28) becomes:

$$SS_{X,i} = \frac{G_i K_i + G_i(F_{1,i} + F_{2,i})s^2}{1 + G_i K_i H_i} \quad (29)$$

From equation (10) and equation (29), the transfer function is

$$SS_{X,i} = \frac{2 + G_i K_i H_i}{H_i(1 + G_i K_i H_i)} \quad (30)$$

Substituting for $G_i(s)$, $H_i(s)$, and $K_i(s)$ from Equations (3), (7), and (13), respectively, into the string stability condition (23), the inequality yields:

$$\left|\frac{(2 + \omega_{K,i}h_d)s^2 + \omega_{K,i}(\omega_{K,i}h_d + 1)s + \omega_{K,i}^2}{(1 + \omega_{K,i}h_d)s^2 + \omega_{K,i}(\omega_{K,i}h_d + 1)s + \omega_{K,i}^2}\right| \leq |1 + h_d\omega_{K,i}s| \quad (31)$$

Substituting $s = j\omega$ into this inequality; we obtain:

$$\frac{\omega^2}{\left(\omega_{K,i}^2 - (\omega_{K,i}h_d + 1)\omega^2\right)^2 + \omega^2\omega_{K,i}^2(\omega_{K,i}h_d + 1)^2} \leq h_d^2 \quad (32)$$

As $\left(\omega_{K,i}^2 - (\omega_{K,i}h_{d,i} + 1)\omega^2\right)^2 > 0$, the inequality can be reduced to:

$$\frac{1}{\omega_{K,i}^2(\omega_{K,i}h_d + 1)^2} \leq h_d^2 \quad (33)$$

which can be solved via $(\omega_{K,i}h_d)^2 + \omega_{K,i}h_d - 1 \geq 0$.

Hence, the string stability region of the controller cut-off frequency and headway time is:

$$\omega_{K,i}h_d \geq (-1 + \sqrt{5})/2 \quad (34)$$

*2) CACC2 Case:* When $\alpha = 1$ and $\beta = 0$, the head-to-tail string stability transfer function is represented as:

$$SS_{X,i} = \frac{X_i}{X_0} = (\Lambda_{CACC,i-1} + \Lambda_{ACC,i-1})\frac{X_{i-1}}{X_0} \quad (35)$$

As we require the head-to-tail transfer function to be string stable, the worst-case scenario is $X_{i-1}/X_0 = 1$, indicating marginal string stability. Hence, by setting $X_{i-1}/X_0 = 1$, we can obtain a more conservative string stability transfer function:

$$SS_{X,i} = \frac{G_i K_i + G_i F_{1,i} s^2}{1 + G_i K_i H_i} \quad (36)$$

Using Equation (10), the string stability transfer function can be reduced to:

---

[1]Marginally string stable means the traffic oscillation is neither amplified nor dampened when it propagates in traffic flow. Here, the marginally string stable between vehicle $i$ and vehicle 0 indicates the position output of vehicle $i$ equals to the position output of the leading vehicle 0, i.e. $X_i/X_0=1$.

$$SS_{X,i} = \frac{G_i K_i + \frac{1}{H_i}}{1 + G_i K_i H_i} = \frac{1}{H_i} = \frac{1}{1 + h_d s} \quad (37)$$

When the headway time $h_d > 0$ and $SS_{X,i} < 1$, the string stability requirement in Equation (23) can be guaranteed.

*3) CACC3 Case:* When $\alpha = 0$ and $\beta = 1$, the head-to-tail string stability transfer function is represented as:

$$SS_{X,i} = \frac{X_i}{X_0} = \Lambda_{CACC,i-2} \frac{X_{i-2}}{X_0} + \Lambda_{ACC,i-1} \frac{X_{i-1}}{X_0} \quad (38)$$

By using the same relaxation approach (marginal head-to-tail string stability) as in CACC2, we can set both $\frac{X_{i-1}}{X_0}$ and $\frac{X_{i-2}}{X_0}$ equal to one, yielding the same string stability transfer function as the second scenario:

$$SS_{X,i} = \frac{G_i K_i + G_i F_{2,i} s^2}{1 + G_i K_i H_i} \quad (39)$$

Thus, we obtain the same string stability requirement in Equation (23) as the second scenario: $h_d > 0$.

*4) ACC Case:* Using a similar method, when $\alpha = 0$ and $\beta = 0$, the head-to-tail string stability transfer function will degrade to:

$$SS_{X,i} = \frac{X_i}{X_0} = \Lambda_{ACC,i-1} \frac{X_{i-1}}{X_0} = \frac{G_i K_i}{1 + G_i K_i H_i}$$
$$= \frac{\omega_{K,i} s + \omega_{K,i}^2}{(1 + \omega_{K,i} h_d) s^2 + \omega_{K,i}(1 + \omega_{K,i} h_d) s + \omega_{K,i}^2} \quad (40)$$

Consequently, by substituting $s = j\omega$, the string stability requirement in Equation (23) yields:

$$\frac{\omega_{K,i}^2 (2 - \omega_{K,i}^2 h_d^2)}{(1 + \omega_{K,i} h_d)^2} \leq \omega^2 \quad (41)$$

Since $\min_{\omega \geq 0} \omega^2 = 0$, this inequality can be solved by:

$$\frac{\omega_{K,i}^2 (2 - \omega_{K,i}^2 h_d^2)}{(1 + \omega_{K,i} h_d)^2} \leq 0 \quad (42)$$

Hence, the string stability region of the controller cut-off frequency and headway time is:

$$\omega_{K,i} h_d \geq \sqrt{2} \quad (43)$$

**Remark:** Based on the above analysis, the decision-making process can be summed up as: (i) positive $\omega_{K,i}$ and $h_d$ to ensure individual vehicle stability; (ii) relatively large $\omega_{K,i}$ to minimize spacing error; (iii) $\omega_{K,i} h_d$ has specific upper-bound for mitigating noise effects; (iv) $\omega_{K,i} h_d$ lies within feasible region to avoid actuator saturation issue; and (v) string stability: $\omega_{K,i} h_d \geq \frac{-1+\sqrt{5}}{2}$ for CACC1, $h_d > 0$ for CACC2 and CACC3, $\omega_{K,i} h_d \geq \sqrt{2}$ for ACC scenario.

## V. NUMERICAL EXPERIMENTS

### A. Numerical experiment design and control parameters

This section presents two numerical experiments to verify the performance of the proposed platoon control method. The first experiment seeks to verify the effectiveness of the proposed CACC-DIFT design in a dynamic IFT environment. The second experiment compares the performance of CACC-DIFT with a CACC with a fixed IFT (CACC-FIFT). CACC-FIFT will reduce to ACC if any of the V2V communications fail because CACC-FIFT is designed for the specific fixed IFT that is assumed to characterize the platoon control.

Consider a 10-CAV platoon with one leading CAV ($i = 0$) and 9 following CAVs. In all experiments, the movement of the leading CAV is predetermined according to NGSIM field data. It contains 240-seconds of vehicle trajectory data collected on eastbound I-80 in the San Francisco Bay area at Emeryville, California. In CACC-DIFT, the first following CAV ($i = 1$) can receive information only from one proceeding vehicle ($i = 0$); so, the controller will switch between CACC2 and ACC if the IFT changes. For the other CAVs ($i = 2, ..., 9$), the controller can switch among the four controllers (i.e. CACC1, CACC2, CACC3 and ACC). To measure the IFT dynamics, a statistical model in [14], which is a function of distance, is used to describe the success rate of a packet delivery between two CAVs.

The desired time headways in all controllers are set to 1 second to prevent traffic oscillation generated by controller switching. Otherwise, the four controllers will have the different stable states, which will lead the additional traffic oscillation during the switching of stable states. The inputs of the parameters of each controller are shown in Table 1. The initial state of the platoon is $v_i(0) = 25 m/s$ for all $i$; the spacing between adjacent CAVs is $x_{i-1}(0) - x_i(0) = h_d v_i(0) + L_i = 30m$, where $L_i$ is the length of CAV $i$ and set as 5m for all the vehicle. In the following experiments, the time interval $T = 0.1s$.

TABLE 1. CONTROL PARAMETERS

| Controller | $w_k$ | $h_d$ | $\alpha$ | $\beta$ |
|---|---|---|---|---|
| CACC1 | 0.8 | | 1 | 1 |
| CACC2 | 0.8 | 1 | 1 | 0 |
| CACC3 | 0.9 | | 0 | 1 |
| ACC | 1.45 | | 0 | 0 |

We compare the performance of CACC-DIFT with that of the benchmark CACC-FIFT. The platoon controller will degenerate to ACC if any V2V communication fails.

### B. Experiment results

The first experiment evaluates the effectiveness of the proposed CACC-DIFT. Specifically, this experiment tests the stability of the CAV controlled CACC-DIFT in the context of time-varying IFT. Figs. 4(a) and 4(b) show the results of the spacing and speed tracking error, respectively.

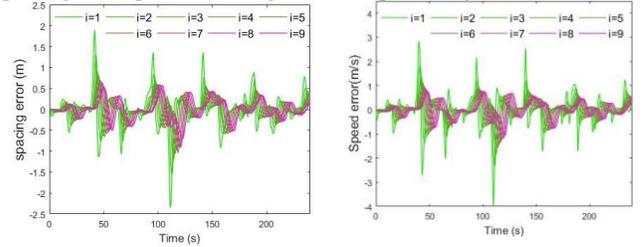

(a). Spacing tracking error     (b). Speed tracking error

Figure 4. Performance of the proposed CACC-DIFT

Fig. 4(a) shows that spacing tracking error between two adjacent CAVs in the CAV platoon reduces according to the order of their positions in the platoon. For example, the maximum spacing tracking error of the first following CAV ($i = 1$) is 2.36m, compared to 1.28m for the second following CAV ($i = 2$) and 0.67m for the last CAV. Fig. 4(b) illustrates that the speed tracking error varies smoothly near the platoon tail. Hence, though the IFT is uncertain, CACC-DIFT maintains good string stability performance.

The second experiment compares the proposed CACC-DIFT with the CACC-FIFT. The performance of these two control schemes is illustrated in Figs. 5(a) and 5(b). Fig.5 (a) shows that the standard deviation of spacing tracking error decreases sequentially in the platoon for both CACC-DIFT and CACC-FIFT. However, CACC-DIFT outperforms CACC-FIFT in that the spacing error of CACC-DIFT reduces faster than that of CACC-FIFT. For example, the standard deviation of spacing tracking error for the last CAV under CACC-DIFT is 0.246m, compared with 0.349m for CACC-FIFT. A similar trend can be found in Fig. 5(b) which shows the standard deviation of the speed tracking error.

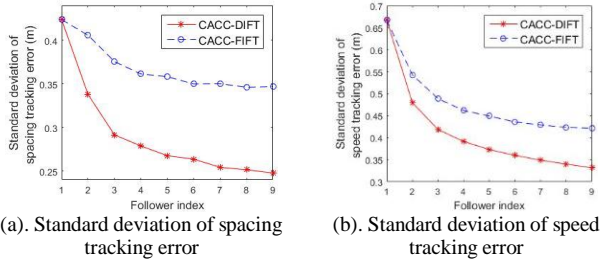

(a). Standard deviation of spacing tracking error   (b). Standard deviation of speed tracking error

Figure 5. Comparsion of CACC-DIFT and CACC-FIFT

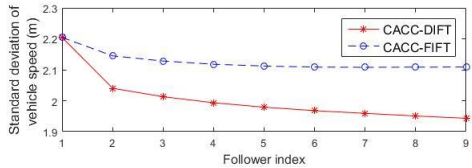

Figure 6. Standard deviation of speed fluctuations

To further investigate the performance benefits under the CACC-DIFT, the performance of the two controller schemes is compared when traffic oscillates (e.g., stop-and-go or slow-and-fast traffic). The standard deviations of the vehicle speed for CACC-DIFT and CACC-FIFT are shown in Fig. 6. It illustrates that the fluctuations in standard deviation of speed decrease under CACC-DIFT as we approach the tail of the platoon, which implies that traffic oscillations are dampened. However, CACC-FIFT cannot reduce the speed fluctuations significantly due to the dynamic nature of the IFT, where CACC will degenerate to ACC if any V2V communications fail. Thereby, CACC-FIFT cannot dampen traffic oscillations efficiently. Based on these results, we conclude that the performance of the CAV platoon controlled by the proposed CACC-DIFT is better and more robust than that of CACC-FIFT in a realistic V2V communication environment.

## VI. Concluding Comments

This study introduces the CACC-DIFT design for CAV platoons by factoring the time-varying nature of the information flow topology arising from V2V communication failures in real-world connected vehicle environments. The CACC-DIFT design is developed for the two-predecessor-following IFT. Four switchable PD controllers are provided to control the CAV platoon under the four possible IFTs by considering the stability of both individual CAVs and the whole platoon. Insights from numerical experiments indicate that compared to the commonly-proposed CACC-FIFT design, the proposed CACC-DIFT mechanism can reduce the spacing tracking and speed tracking errors, and dampen traffic oscillation much faster. Thereby, CACC-DIFT is string stable and outperforms CACC-FIFT considerably.